# Graphene-based Bolometers

Xu Du[1], Daniel E. Prober[2], Heli Vora[1], Chris Mckitterick[2]

[1]Department of Physics and Astronomy, Stony Brook University

[2]Departments of Applied Physics and Physics, Yale University

I.   Bolometer and device principles
II.  Advantages and challenges of graphene-based bolometers
    2.1.   Electronic and transport properties of graphene
    2.2.   Electron-phonon scattering and phonon cooling
III. Graphene-based bolometers: approaches
    3.1.   Semiconducting bilayer graphene
    3.2.   Johnson noise thermometry
    3.3.   Superconductor-graphene junction

IV.  Conclusions and future challenges

## I.   Bolometer and device principles

Modern photon detectors are widely employed in sensitive applications ranging from astrophysical observations to quantum communications[1-4]. The configurations considered to date for ultrasensitive Terahertz graphene detectors are thermal detectors at low temperatures[5-8], typically $T \leq 1$ K[7,9-12]. In this review article we consider such graphene photodetectors for the far-infrared (Terahertz) frequency range. Graphene detectors have been investigated with higher power signals from far-infrared to mid-infrared[13,14] and in the near-IR to optical range[15,16]. These other applications utilize different detection modes for optimum sensitivity, including the photo-thermal-electric effect, due to the much larger power or photon energy.

We consider a potential application for detecting THz photons: operation in a cold, space-based observatory.[2,17]  From 1 THz to 10 THz there are sensitive molecular spectroscopy applications in astrophysics that have background count rates of order 100 counts per sec.  This assumes a cold, tunable bandpass prefilter with Q = 1,000.  A spectral line signal from a distant molecular source might have a count rate after the bandpass filter of order 1 to 100 times the background, up to $\approx 10^4$  counts/sec.  The designs we discuss would serve in this application.

Various modes of readout have been explored for the sensitive graphene THz detectors, including measurement of Johnson noise[6,10], of superconducting critical current[11,12], and of the resistance of superconducting tunnel contacts[9]. To define the challenges associated with photon detection with graphene, in this section we outline the detector concepts and some of the graphene-related issues that are known, or that require further research. We then summarize the important performance metrics.

An energy detector (calorimeter) works by absorbing a photon and reading out the resulting temperature increase. The schematic device structure of a thermal detector is shown in Figure 1. We show the detector element at the center of a planar antenna; various antenna designs can be used. An antenna is needed because best sensitivity is achieved with a small thermal detector, of micron size scale or smaller, whereas the photon wavelength is 100s of microns. The antenna allows efficient coupling of the photon to the small detector. Thermal single-photon detectors based on the superconductor transition provide good models of thermal detectors. These have been explored for energies $\geq 0.15$ eV[17,18] and studied extensively in the near-IR/visible range[4] and in the x-ray range.[19]

A thermal detector can also be used to measure power, in which case it is called a bolometer. In the discussion below we will call both kinds of detectors 'bolometers'. The bolometer as power detector measures the difference in power absorbed with the incoming beam on and off, as in Fig. 1. For linear operation the response time is set, as in the calorimeter, by the specific heat C and the thermal conductance G; $\tau = C/G$. The thermal conductance is[7]

$$G = G_{eph} + G_{diff} + G_{photon} \qquad (1)$$

$$G = dP/dT$$

$G_{eph}$ is due to emission of phonons by the heated electrons, $G_{diff}$ is due to cooling of the electrons by diffusion out into the colder contacts at a bath temperature $T_o$ [20,21], and $G_{photon}$ is due to emission of microwave photons (Johnson noise) which remove energy from the detector until it returns to the quiescent temperature, $T_o$. Generally, the best sensitivity for both energy and power detection is achieved by minimizing G, to achieve a long measurement time and large power response. Diffusion cooling of the graphene can be minimized by use of superconducting contacts[17,22,23], discussed later. Photon cooling is unavoidable with the Johnson noise readout method[7] since one wishes to maximize the Johnson noise 'signal'. For cooling by emission of phonons, the power emitted, in one of the most important regimes (discussed below in Sec. 2.2) is given as

$$P_{eph} = \Sigma A (T^4 - T_o^4)$$

$$G_{eph} = 4 \Sigma A T^3$$

$\Sigma$ is the electron-phonon coupling constant and A is the device area[7].

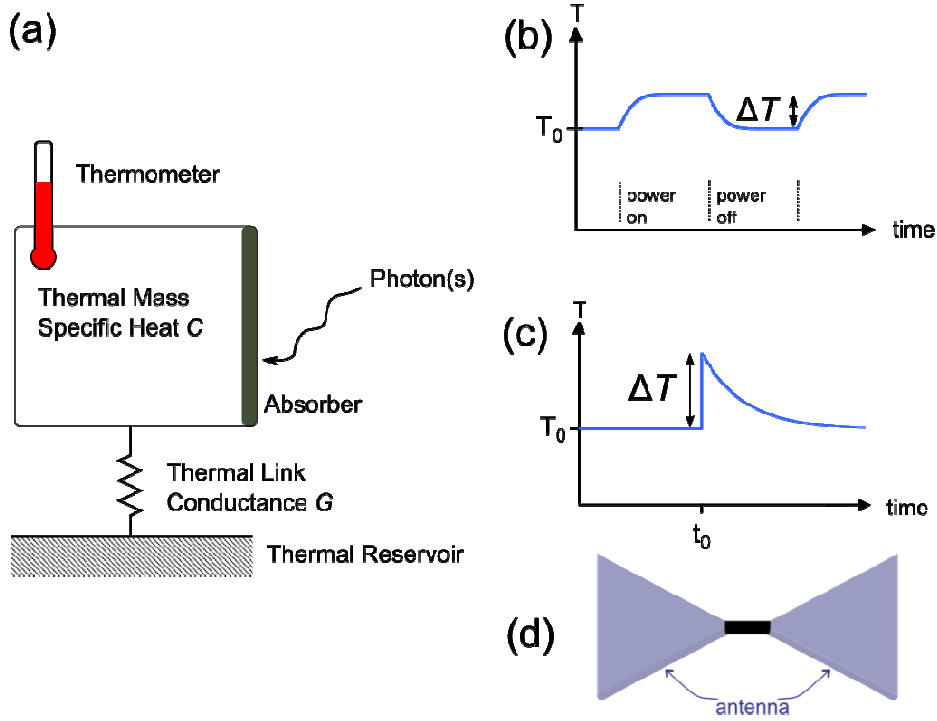

Figure. 1 (a) Schematic of bolometer or calorimeter. C is the heat capacity and G is the thermal conductance. In a graphene detector, C and the thermometer function are provided by the electron subsystem. G has contributions from diffusion to the contacts ($G_{diff}$), phonon emission ($G_{eph}$) and Johnson noise emission ($G_{photon}$). (b) Temperature response, T(t), for a bolometer and (c) for a single-photon calorimeter. For graphene, the resistance is nearly independent of temperature. (d) Schematic of antenna coupling to a small graphene device at the center. For a THz detector, the antenna linear dimension is a few 100 µm. The contacts for readout at microwave or lower frequency are not shown.

The single-photon detector responds with an initial temperature increase of

$$\Delta T = E/C \tag{2}$$

if $\Delta T \ll T_0$, the linear range of photon detection. We use the notation $\Delta$ to indicate signal changes, and $\delta$ to indicate the distribution widths. The figure of merit for the energy detector is the energy Resolving Power, $R = E / \delta E$, with E the photon energy and $\delta E$ the energy width. The usual convention is to list the full-width at half-maximum (FWHM) of the distribution of measured photon energies when illuminating with single photons of fixed energy. In this paper we shall designate $\delta E = \delta E_{FWHM} \approx 2.35\ \delta E_{rms}$, with $\delta E_{rms}$ the root-mean-square (*rms*) energy width. The total energy width is

$$\delta E = \left(\delta E_{int}^2 + \delta E_{readout}^2\right)^{1/2} \tag{3}$$

with intrinsic energy resolution $\delta E_{int}$ and readout energy resolution $\delta E_{readout}$ defined below.

For the bolometer power detector, the response is $\Delta T_{power} = \Delta P/G$ for linear response. The main figure of merit is the noise-equivalent-power, NEP[7,24], the minimum power that can be detected with a 1-Hz output readout bandwidth, in $W/(Hz)^{1/2}$.

We will treat three specific detector designs in Table 1 to frame the later discussions. We consider the photon energy for a 1 THz photon, $E = 7 \times 10^{-22}$ J. The two specific device designs are chosen based on optimizations done previously.[7] The first, A, has small area and is designed to achieve good energy resolution for detecting single photons using a temperature readout that measures the Johnson noise. For device A, the thermal response $\Delta T$ is not linear in single-photon energy in the THz range. (See Fig. 9a below.)

The other designs have a much larger graphene area, chosen to allow operation in the linear range, with $\Delta T \ll T_o$. Two readout methods are considered: Design B, using Johnson noise readout with an ultralow-noise amplifier with 150-MHz bandwidth[7], and Design C, measurement of the temperature dependent resistance of superconductor tunnel junction contacts. The large area of designs B and C is impractical for single photon detection. Design A does not respond linearly to photon energy, so is not considered in the table for power detection, though power measurement with Design A can be accomplished by photon counting.[2] The device parameters scale with temperature as $C \propto T$, $G_{eph} \propto T^3$ (within one model; see Sec. 2.2) and for these $\tau_{eph}^{-1} = C/G \propto 1/T^2$. If we extrapolate the measured value[6] of $G_{eph}$ from higher temperatures,[6] we find $\tau_{eph}$ = 45 µs at 0.1 K.[7] We use this value for the results in Table I and Figure 9. For design A, the effective time constant during the pulse is 0.5 µsec, and the average temperature increase during the pulse is 0.5 K.

The intrinsic energy resolution due solely to device thermodynamic fluctuations is[25]

$$\delta E_{int} = 2.3\sqrt{k_B T^2 C} \tag{4}$$

with $k_B$ being Boltzmann's constant. This formula applies when the amplifier and bias circuits do not affect the thermal properties of the detector[26]. (If, instead, there is negative electrothermal feedback-ETF then the prediction[26] for $\delta E_{int}$ is less than Eq. 4.) We use Eq. 4 in our discussion because the effects of ETF should be small when using the Johnson noise readout or the resistive readout of Design C. For detectors that have a linear response to photon energy, such as design B, the predicted total energy resolving power is $R \leq 1$ for all practical values of parameters at 0.1 K. One needs energy resolving power $R > 3$ for a practical photon counter with linear response, for which $\Delta T \ll T_o$. Going to lower temperatures would improve the energy resolution of design B only slightly, since C needs to <u>increase</u> as $1/T_o$ to stay in the linear range of operation. Moreover, maximum count rates would be much too small for $T_o \ll 0.1$ K. For design A, the energy width is $\delta E_{FWHM}$ = 0.45 E for photon detection at 1 THz, but $\delta E_{FWHM}$ = 0.2 E for sampling the baseline, when no photons are present. This provides good enough separation to allow photon counting with design A, as we discuss below with Fig. 9. We conclude that operation in the non-linear range is necessary for THz single photon detection.[7]

| Design | Area ($\mu m^2$) | $C(T_o)$ (J/K) | Energy Resolving Power $R = E/\delta E_{FWHM}$ | $NEP_{tot}$ (W/Hz$^{1/2}$) | Operation |
|---|---|---|---|---|---|
| A | 4.5 | $3 \times 10^{-22}$ | 2.2* | - | Non-linear in photon energy. Johnson noise readout |
| B | 1000 | $7 \times 10^{-20}$ | 0.5 | $1.2 \times 10^{-19}$ | For count rates to $\approx 10^5$ /sec. Johnson noise readout |
| C | 1000 | $7 \times 10^{-20}$ | - | $5 \times 10^{-20}$ | For count rates to $\approx 10^4$ /sec. SC tunnel junction resistance-readout of R(T) |

Table 1. Model device design parameters, and predicted performance[7]. For these designs, we consider an electron density n = $10^{12}$ /cm$^2$, T = 0.1 K, and a THz photon $E_{ph}$ = 7 x $10^{-22}$ J. *See discussion in the text of required value of *R* for good photon counting in the non-linear range of operation. For device A, the thermal response for different photon energies is not linear, as seen in Fig. 9 and discussed in Sec. IV; the thermal response time is 0.5 µsec for a 1-THz photon.[7] For designs B and C, the response is reasonably linear in power for count rates up to $\approx 10^5$ /sec for design B and up to $\approx 10^4$ /sec for design C. The electron-phonon time is 45 µsec at $T_o$ = 0.1 K.

We next consider an overview of the performance of a bolometer power detector based on graphene. The total NEP includes the intrinsic thermodynamic energy fluctuations, which lead to Eq. 3, and the uncertainty of determining the exact temperature when one reads out the temperature via Johnson noise emission.

$$NEP_{tot} = \left(NEP_{int}^2 + NEP_{readout}^2\right)^{1/2} \qquad (5)$$

The formula for the intrinsic NEP (FWHM) is $NEP_{int} = 2.3\sqrt{4k_B T_0^2 G(T_0)}$ for device parameters that give linear operation detecting individual photons[24] such as designs B and C. As discussed for design C later in this article, the detector can be designed so that $NEP_{readout}$ is small compared to the intrinsic NEP. Design A has non-linear response of ΔT as a function of single-photon energy, so a more complex calculation of NEP would be required. The measurement of power with design B will have reasonably linear response for photon arrival rates $\leq 10^5$/s. At higher count rates there will be saturation and larger noise (larger NEP) than indicated in the table. For those larger count rates, photon shot noise[24] will dominate the NEP. We note that the photon counting mode of design A is usually a better choice for measuring power for nearly monochromatic photons[2].

For design C, where graphene-superconductor tunnel junction resistance is used as readout, a slightly lower NEP can be achieved, as discussed in section 3.3. However, since the R(T) dependence is highly non-linear, the photon arrival rate is limited to a lower value of ~$10^4$/s (see section 3.3) We summarize in Table 2 the desired system and engineering characteristics that a THz detector should meet.

|   | **Desired Device Characteristic** | **Desired Parameter** |
|---|---|---|
| 1 | Impedance at THz match planar antenna | $\approx$ 100 ohms, possibly use plasmonic coupling[27] |
| 2 | Impedance match circuit at readout frequency, typically 0.1 – 1 GHz | $\approx$ 50 ohms |
| 3 | Lateral dimensions between antenna terminals is small, for low stray impedances of the antenna structure at THz | Few μm between antenna terminals. Graphene that extends outside the antenna terminal contacts is allowed. |
| 4 | Substrate and gate electrode not attenuate or interfere at THz and readout frequency | Insulating Si or good insulator; small metal gate |

Table 2. Desired Characteristics of a Graphene THz Photon Detector. The power coupling efficiency of the graphene to an rf/microwave amplifier, or the photon coupling efficiency of the THz antenna to the graphene, are each given by the formula, Eff = $4 (R_A R_g)/(R_A + R_g)^2$, with $R_A$ the amplifier or antenna impedance (typically 50 ohms and $\approx$ 100 ohms, respectively), and $R_g$ is the resistance of the graphene. When $R_g = R_A$ the coupling efficiency is unity.

**Advantages and Challenges of graphene-based bolometers**

There are several advantages in utilizing graphene for bolometers applications. Graphene, as a single atomic layer of carbon, has ultra-small volume. At the same time, from its Dirac fermion electronic structure, graphene has low electron density of states. As a result the material has low heat capacity. This allows large intrinsic energy resolving power for single photon detection and fast device response. Additionally, the electron-phonon interaction in graphene is weak at low temperatures, as a result of the small Fermi surface. This allows a very small electron-phonon thermal conductivity and therefore high intrinsic sensitivity for graphene-based bolometers if phonon cooling is dominant. Finally, graphene, as a 2D material, allows relatively low device resistance compared to one-dimensional nanomaterials (e.g., carbon nanotubes[28]). The low and gate-tunable resistance makes it possible to integrate the devices with a planar antenna with high coupling efficiency.

Along with the above advantages are technical challenges. A major challenge is that, with weak electron-phonon scattering, the resistance is only weakly temperature dependent. It is thus challenging to measure the electron temperature change due to incoming radiation power. In addition, it is challenging to thermally isolate graphene in order to achieve the small electron-phonon thermal conductance. One must also design the devices to have low microwave impedance, in order to match with the antenna and external microwave readout circuit. In the following sections, we will discuss in detail these advantage and challenges.

**2.1. Electronic and transport properties of graphene**

The unique electronic properties of graphene originate from its lattice symmetry and the atomic structure of carbon[29]. Graphene is a single atomic layer of sp2-bonded carbon atoms arranged in a honeycomb crystal lattice. Each unit cell of this honeycomb lattice has two carbon atoms. The 4th valence electron half-occupies the $p_z$ orbital which extends perpendicular to the graphene plane. The side-ways overlap of these orbitals forms the weak π-bonds, which determine the electrical conductivity of graphene.

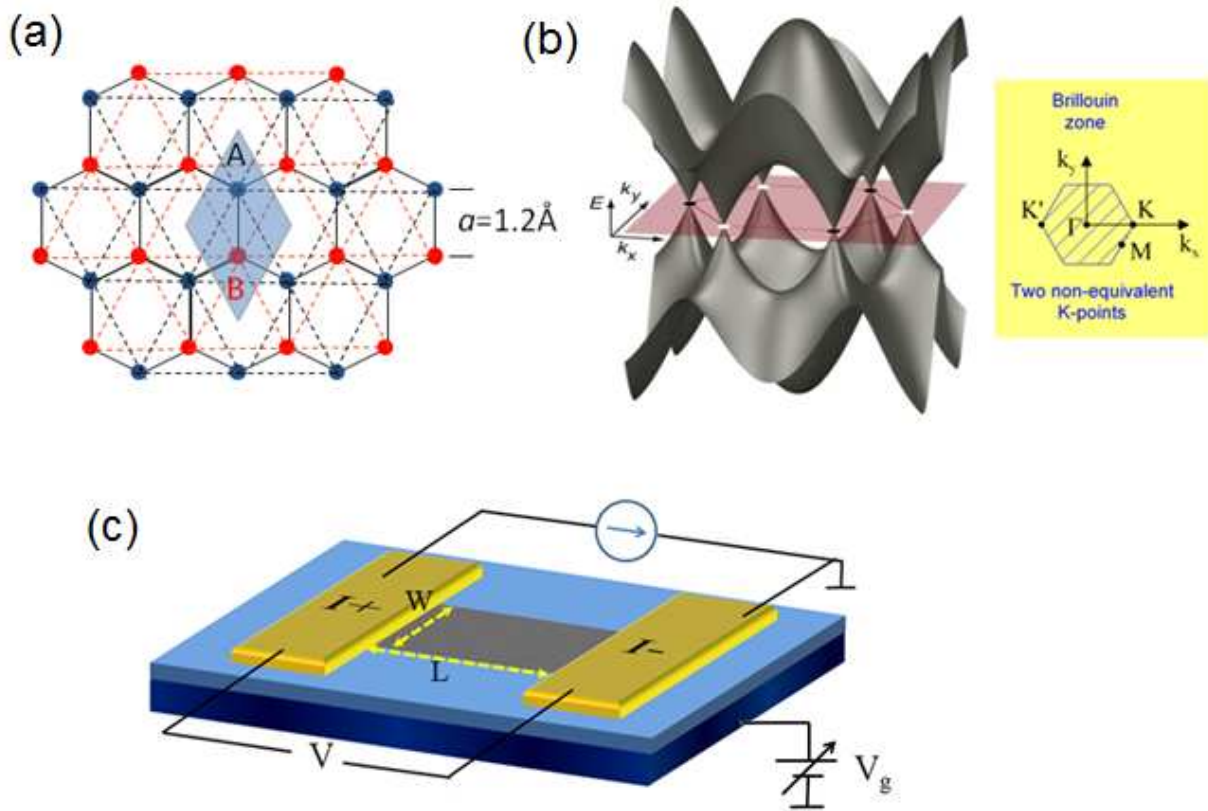

Figure 2. (a) Atomic structure of graphene. The honeycomb lattice is an overlay of two sets of triangular lattices with inversion symmetry. Each unit cell of the honeycomb lattice has two atomic sites (A and B). (b) Energy dispersion of graphene, from ref[30]. The conduction and valence bands touch at Dirac points (K and K'), near which the energy dispersion $E(k)$ is approximately linear. (c) Schematics of a graphene field effect device. $W$ and $L$ denote the width and length of graphene between the source and drain electrodes.

The symmetry of the honeycomb lattice gives rise to an energy dispersion shown in Figure 2. In intrinsic graphene the Fermi level aligns with the K and K' points where conductance and valence bands meet. The K and K' points are the so-called Dirac points, near which the energy dispersion is linearly conical corresponding to a Hamiltonian

$$\hat{H} = \pm v_F \vec{\sigma} \bullet \vec{p}.  \tag{6}$$

where $v_F \approx 10^6 \, m/s$ is the Fermi velocity, $\vec{\sigma}$ are the Pauli matrices, and $\vec{p}$ is the momentum relative to the Dirac point. Since there are two non-identical carbon atoms per unit cell, the wave functions have the form of a spinor $\begin{pmatrix} \psi_A \\ \psi_B \end{pmatrix}$, where A and B denote the two atomic sites. This gives rise to an additional degree of freedom, the pseudospin, which describes the distribution of the wavefunction on the two atomic sites. The pseudospin vector $\vec{\sigma}$ is either parallel or anti-parallel to the momentum, and $\vec{\sigma} \bullet \hat{p} = \pm 1$ ($\hat{p}$ being the unit vector of momentum) gives the chirality of the electronic excitations, the quasiparticles.

As a result of the linear energy dispersion, the 2D electron density of states (DOS) has a linear energy dependence in graphene: $N(E) = \frac{2E}{\pi(\hbar v_F)^2}$, where $\hbar$ is the reduced Planck constant. The DOS approaches zero at the charge neutral Dirac points. Due to its small volume and low DOS, graphene has very small electron heat capacity. Considering the simple case of an electron gas in which $C_e = A \int \varepsilon N(\varepsilon) \frac{df(\varepsilon)}{dT} d\varepsilon$ (here $C_e$ is the electron heat capacity, A is the area of graphene, $N(E) = \frac{2E}{\pi(\hbar v_F)^2}$ is the DOS in graphene, and $f(\varepsilon)$ is the Fermi distribution function), one can estimate the value of $C_e$ in graphene, and its dependence on gate voltage and temperature, as illustrated in Figure 3. The heat capacity depends linearly on temperature except at the Dirac point, where a $T^2$ dependence is expected. At the technically relevant conditions, we find that the heat capacity in graphene can easily reach extremely small values (e.g., $C_e \sim 10^{-21}$ J/K for T < 5 K at Vg~10 V and n ~ $7 \times 10^{11}$ cm$^{-2}$, for a 1 μm$^2$ sample). This small value cannot be achieved in conventional metal structures. Useful for later discussions in section 3, we note that at $T = 0.1$ K and n = $10^{12}$ cm$^{-2}$, the heat capacity is ~ $7 \times 10^{-23}$ J/K.

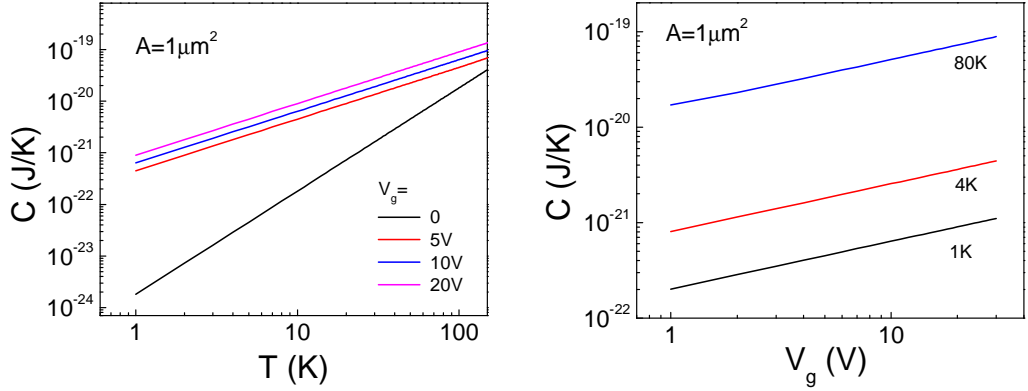

Figure 3. Calculated gate voltage and temperature dependence of the electron heat capacity for a 1μm$^2$ graphene. Here we assume the graphene is on top of a SiO$_2$(300nm)/Si substrate.

The conductivity of graphene can be described by the Boltzmann transport equation:

$$\sigma = \frac{e^2 v_F^2 N(E_F) \tau(k_F)}{2} \tag{7}$$

Here $N(E_F) = \frac{2E_F}{\pi(\hbar v_F)^2}$ is the DOS at the Fermi level, and $\tau(k_F)$ is the scattering time. Different types of charge carrier scattering mechanisms give rise to their scattering rate that depends on Fermi wave vector/energy[31]:

$$\frac{\hbar}{\tau(k_F)} = \frac{n_i^{scatt}}{8} N(E_F) \int d\theta |V_{scatt}(q)|^2 (1 - \cos^2(\theta)) \tag{8}$$

where $n_i^{scatt}$ is the impurity density, $V_{scatt}(q)$ is the Fourier transform of the scattering potential, and $q = 2k_F \sin(\theta/2)$. It is believed that the dominant scattering in graphene is from charged impurities which induce Coulomb scattering[31,32], with scattering time: $\tau_{k_F} \propto \frac{k_F}{n_i^C}$ and correspondingly a conductivity $\sigma \propto \frac{k_F^2}{n_i^C} \propto E_F^2$. Here $n_i^C$ is the density of the charged impurity scattering centers. Short range scattering from point defects and phonons[31,33] also plays an important role in limiting the conductivity of graphene. In contrast to the long range Coulomb scatterers, the short range scatterers give a scattering time $\tau_{k_F} \propto \frac{1}{n_i^S k_F}$ and correspondingly an energy independent conductivity $\sigma \propto \frac{1}{n_i^S}$, where $n_i^S$ is the density of the short range scatterers.

The charge carriers in graphene can also be scattered by vacancies and corrugations[31,34], which form bound states call the mid-gap states. The mid-gap states scattering contribute a scattering time $\tau_{k_F} = \frac{k_F}{\pi^2 v_F n_i}[\ln(k_F R_0)]^2$, and conductivity $\sigma = \frac{2e^2}{\pi h} \frac{k_F^2}{n_i}[\ln(k_F R_0)]^2$. Here $n_i$ and $R_0$ are the density and spatial size of the mid-gap state scatterers.

In a graphene field effect device (Figure 2 c), the carrier density in graphene can be tuned by capacitively inducing charge carriers using a gate voltage: $n = \frac{\varepsilon \varepsilon_0 V_g}{ed}$, where $V_g$ is the gate voltage, and $\varepsilon$ and $d$ are the dielectric constant and the thickness of the gate insulator, respectively. Consequently the Fermi energy and Fermi wave vector can be tuned: $E_F = \hbar v_F \sqrt{n\pi}$ and $k_F = \sqrt{n\pi}$. The experimentally observed gate voltage dependence of the resistivity is a direct result of the tuning of Fermi energy and scattering time. For example, the Coulomb scattering contributes resistivity which has a $1/V_g$ dependence, while the short range scatterers contribute to a gate-voltage-independent resistivity. The combined effect, summed up using Matthiessen's rule, gives the commonly observed R-$V_g$ dependence[35].

### 1.2 Electron-phonon scattering and phonon cooling

Phonon emission due to electron-phonon scattering ultimately limits the intrinsic thermal conductivity of the hot electrons in graphene, and thus determines the sensitivity of the graphene-based bolometers. The linear electronic dispersion and 2D nature of electrons as well as phonons in graphene lead to a unique electron-phonon interaction compared to what has been found in conventional metals, semiconductors and 2DEG systems.

Under different experimental conditions, different phonon modes may contribute to the electron-phonon scattering. The optical phonon energy in graphene ($\omega_0$) is about 200meV[36,37] which is well above the operating temperature range of graphene devices so far designed for sensitive bolometric detection. Bolometric detection devices that we focus on have graphene sitting on a substrate. Therefore, we do not treat the out-of plane flexural phonons, which might play a role in heat conduction in the case of suspended graphene[38-40]. For graphene on a dielectric substrate, the coupling to the substrate phonons must be considered as substrate optical phonons might be excited if the device operation temperature is high, and therefore these excitations would contribute to cooling. Depending on the choice of substrate, this might be important. However, we limit our discussion to the devices fabricated on $SiO_2$ substrates. Here it had been demonstrated that the substrate "remote" phonon contribution to electron-phonon scattering is only evident for T > 200 K[41].

The most important contribution to the electron-phonon scattering is from acoustic phonons in graphene. The coupling constants for transverse acoustic phonons (TA) are about an order of magnitude smaller than those for longitudinal acoustic (LA) phonons, and the TA phonons do not contribute significantly to heat conduction[42]. We therefore discussion electron – LA phonon coupling.

Most theoretical studies of the graphene electron-phonon interaction assume that an average electronic temperature can be defined. That is, heat is distributed in the electron system via electron-electron interactions much faster than it is given off to lattice, and the occupation of energy levels is given by the Fermi function at an effective electron temperature. This assumption is justified since electron-electron interactions have a much shorter time scale than electron-phonon interactions (e.g., femto-second vs. pico-second, at room temperature[43,44]).

For pure graphene, two temperature ranges are important in studying the electron-phonon interaction: a high temperature range with $T \gg T_{BG}$, also termed the equipartition (EP) regime; and a low temperature range with $T \ll T_{BG}$ called Bloch-Gruneisen (BG) regime. Here, $T_{BG}$ is the Bloch-Gruneisen temperature given by

$$k_B T_{BG} = (2s/v_F) E_F \approx 0.04 E_F \tag{9}$$

where $s$ is the sound velocity of $2 \times 10^4$ m/s in graphene. At a carrier density of $n \sim 10^{12} cm^{-2}$ $E_F = \hbar v_F \sqrt{n\pi} \sim 1.8 \times 10^{-20} J$, corresponding to a $T_{BG} \sim 50K$. Electron-phonon interactions also limit the intrinsic carrier mobility in graphene when no impurity scattering is present. This was studied with measurements of the resistivity vs. temperature to determine the transport scattering time $\tau_{trans}$[35,45,46]. The temperature dependence was determined to be $\tau_{trans}^{-1} \sim T^4$ in the low temperature (BG) regime and $\tau_{trans}^{-1} \sim T$ in the high temperature (EP) regime. In the EP regime all phonon modes are populated equally and in the BG regime, a bosonic distribution of the phonons applies. The temperature dependence of the resistivity at different carrier densities is shown in Figure 4.

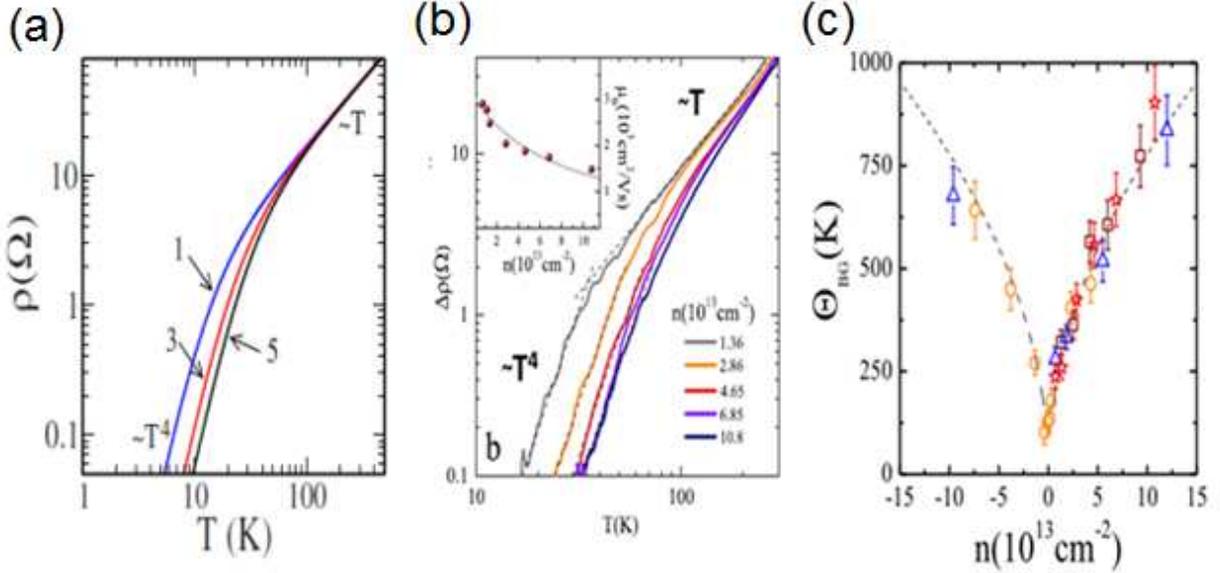

Figure 4. Effect of electron-phonon scattering on transport. (a) Calculated temperature dependence of resistivity $\rho$ at $1 \times 10^{12}$, $3 \times 10^{12}$ and $5 \times 10^{12}$ cm$^{-2}$ in the low temperature (BG) regime (T$^4$) and high temperature (EP) regime (T). From Hawng E. et al.[45]. (b) Experimental results from Efetove et.al.[46] showing the temperature dependent part of the resistivity $\Delta\rho(T)$ from electron-phonon scattering. Inset shows the mobility $\mu_0$ at T = 2 K as a function of the density, which fit (grey line) with combined short and long range impurity scattering. (c) Carrier density dependence of the BG temperature (here $\Theta_{BG} \equiv T_{BG}$).

The T$^4$ temperature dependence of the resistivity in the Bloch-Gruneisen regime was confirmed at extremely high carrier densities using an electrolytic gate[46]. In the low and moderate density regime relevant to the SiO$_2$-based photon detectors, electron-phonon scattering gives a negligibly small contribution to the resistivity compared to other scattering mechanisms[36,45]. It is therefore useful to study electron-phonon scattering through the electron-phonon thermal conductivity for graphene systems of low and moderate density.

**Clean limit**

In this section, we briefly discuss the temperature dependence of the phonon cooling power without taking disorder into account. Generally this can be calculated from the Boltzmann equation in which the occupation probability of an electron excitation with momentum $\hbar k$ in band α is given by[36,47,48]

$$\partial_t f_k^\alpha = S_{e-ph}(f_k^\alpha) \tag{10}$$

where S is the collision integral given by,

$$S_{e-ph}(f_k^\alpha) = -\sum_{p\beta}[f_k^\alpha(1-f_p^\beta)W_{k\alpha \to p\beta} - f_p^\beta(1-f_k^\alpha)W_{p\beta \to k\alpha}] \tag{11}$$

Using Fermi's golden rule one can calculate scattering rates

$$W_{k\alpha \to p\beta} = 2\pi \sum_q w_q^{\alpha\beta}[(N_q+1)\delta_{k,p+q}\delta(\varepsilon_{kp}^{\alpha\beta}-\omega_q) + N_q \delta_{k,p-q}\delta(\varepsilon_{kp}^{\alpha\beta}+\omega_q)] \tag{12}$$

where $N_q$ is the Bose distribution function of a phonon with wave vector q. The energy exchanged with the phonon heat bath is $\varepsilon_{kp}^{\alpha\beta} = \varepsilon_{k\alpha} - \varepsilon_{p\beta}$. Here, $w_q^{\alpha\beta}$ is the transition matrix element that depends on the coupling mechanism between electron-phonon. In case of acoustic phonons with linear phonon dispersion $\omega_q = sq$ and deformation potential coupling $w_q^{\alpha\beta} = D^2 q^2 (1 + s_{\alpha\beta}\cos\theta)/4\rho_m \omega_q$, where $s_{\alpha\beta} = \pm 1$ for interband (-) and intraband (+) scattering. $\theta = \theta_p - \theta_k$ is the relative angle between incoming and outgoing electron momenta and $\rho_m$ is graphene's mass density. Correspondingly the phonon cooling power is given by:

$$P = -\partial_t \sum_{k\alpha} \varepsilon_{k\alpha} f_k^\alpha = -\sum_{k\alpha} \varepsilon_{k\alpha} S_{ph}(f_k^\alpha) \tag{13}$$

Based on the Boltzmann equation approach, detailed calculations can be carried out and analytical solutions can be obtained at several limits through expansion of P up to leading order in $s/v_F$, taking advantage of the large difference between the two velocities[36,48,47]. It is instructive to note that interband transitions between valence and conduction bands do not contribute significantly to cooling power since they require phonon energy greater than $\hbar v_F q$, which cannot be provided by acoustic phonons having energy $\hbar\omega_q = \hbar s q$. Thus, only intraband scattering contributes significantly to phonon cooling power[37,48].

To obtain an analytical expression, it is necessary to evaluate the cooling power in the limit of highly doped ($\mu \gg k_B T_e$) or neutral graphene ($\mu \ll k_B T_e$), $\mu$ being the chemical potential. In these limits the cooling power in graphene follows the familiar power-law temperature dependence as in higher dimensional material [47,48], with $P = A\sum \left(T_e^\delta - T_{ph}^\delta\right)$. Here, A is the area of the graphene in our case, $\Sigma$ is the coupling constant, $T_e$ is the electronic temperature, and we assume a finite lattice temperature $T_{ph}$. An expression for the highly doped, low temperature regime $T<T_{BG}$ is given by[47,48],

$$P = A\Sigma(T_e^4 - T_{ph}^4), \quad \Sigma = \frac{\pi^2 D^2 |\mu| k_B^4}{15 \rho_m \hbar^5 v_F^3 s^3} \tag{14}$$

Here D is the deformation potential. This $T^4$ power law was confirmed experimentally[6,10], although values found for $\Sigma$ differed by ~$10^2$ in the two experiments.

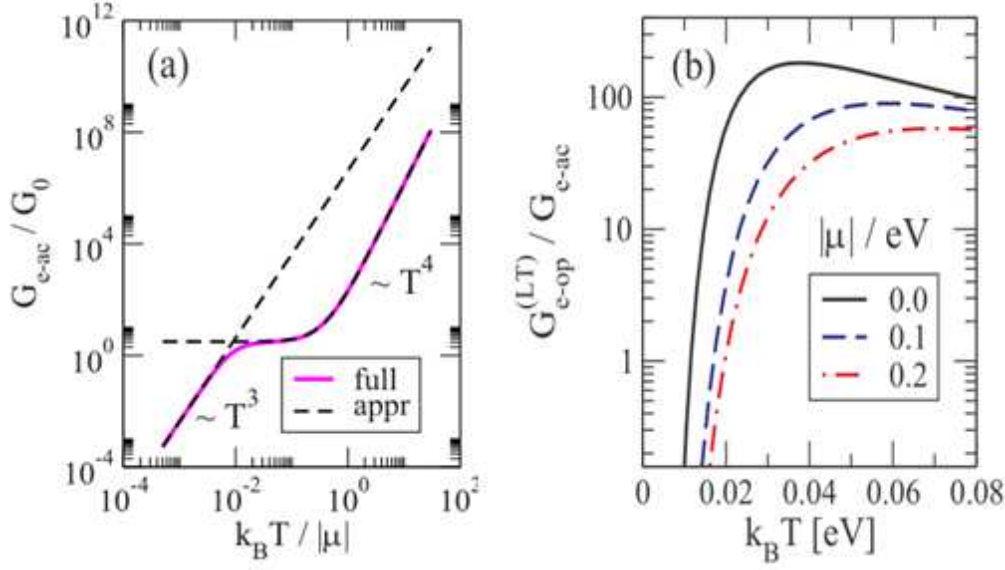

Figure 5. Phonon cooling in single layer graphene in the clean limit, from ref[48]. (a) Temperature dependence of acoustic phonon cooling thermal conductance $G_{e\text{-ac}}$ (normalized by $G_0 = \dfrac{g_e A D^2 \hbar |\mu|^4 k_B}{8\pi^2 \rho_m (\hbar v_F)^6}$, where $g_e$ is the degeneracy). (b) The ratio of thermal conductance between LT optical phonon $G_{e-op}^{(LT)}$ and acoustic phonon $G_{e-ac}$.

In the highly doped, low temperature regime, the energy relaxation time temperature dependence is given as $\tau_{eph} \sim T^{-2}$, different from $\tau_{trans} \sim T^{-4}$ dependence of the transport relaxation time[48]. In the case of the neutral high temperature regime where $\mu \ll k_B T_e$ the temperature dependence becomes much more complicated than the low temperature limit[36,48].

To include effects of screening, the transition matrix elements $w_q^{\rho\beta}$, which depend on the coupling mechanism between electrons and phonons, should be divided by graphene's dielectric function and the cooling power scales as $P \sim T^6$ ($\sigma \sim T^{-5}$)[42,45]. These effects are usually neglected[45], since phonon coupling matrix elements arise from overlap in adjacent atom orbital and are not due to the Coulomb potential, which would be affected by dielectric screening.

**Bilayer Graphene**

In bilayer graphene, the two layers can sense different electrostatic potentials if appropriate gate voltages are applied from above and below the graphene. One can tune the electron filling such that the temperature-dependence of the resistance can be used as an electron thermometer. In this case the resistivity is very large. Thus, for bolometer applications, the coupling efficiency to a planar antenna will be low.

Bolometric detection using bilayer graphene[5] requires understanding of the electron-phonon interaction in this system. Bilayer graphene differs from monolayer graphene by having an approximately parabolic band structure. The cooling power (related to the thermal conductance by $G = \dfrac{dP}{dT}$) in this case of bilayer graphene, below the Bloch Gruneisen temperature of bilayer $k_B T_{BG,BLG} = 2(s/v_F)\sqrt{\gamma_1 |\mu|}$ with $\gamma_1$ band parameter, also scales as $T^4$. The coupling parameter $\Sigma$ in this case is given as:

$$\Sigma = \frac{\pi^2 D^2 \gamma_1 k_B^4}{60 \rho_m \hbar^5 v_F^3 s^3} \sqrt{\frac{\gamma_1}{|\mu|}} \tag{15}$$

Multilayer graphene had also been studied for phonon-limited resistivity[49], which is outside the scope of this article.

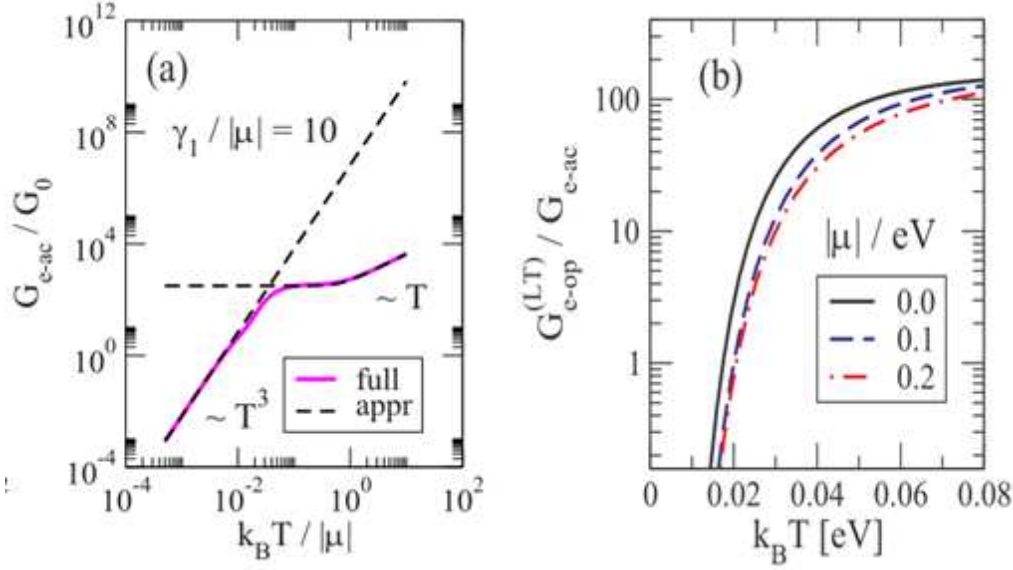

Figure 6. Phonon cooling in bilayer graphene in the clean limit, from ref[48]. (a) Temperature dependence of acoustic phonon cooling thermal conductance $G_{e\text{-ac}}$ (normalized by $G_0 = \dfrac{g_e A D^2 \hbar \gamma_1 |\mu|^3 k_B}{16\pi^2 \rho_m (\hbar v_F)^6}$ ). (b) The ratio of thermal conductance between LT optical phonon $G_{e-op}^{(LT)}$ and acoustic phonon $G_{e-ac}$.

**Effects of disorder**

In the above discussion, disorder and other scattering sources were not taken into account. In most graphene devices where disorder is strong, it has been shown[37,42] that the effects of disorder become important above and below $T_{BG}$ due to different mechanisms.

Due to graphene's small Fermi surface compared to usual metals, the Bloch Gruneisen temperature dictates when quantum effects become important instead of the Debye temperature. $T_{BG}$ is defined by the maximum phonon momentum that can cause a transition for an electron: $q_{max} = 2k_F$. Therefore, above $T_{BG}$ only a fraction of the available phonons, those with $q_{max} \leq 2k_F$, can participate in cooling (see Figure 7); the energy transferred per scattering event is less than $k_B T_{BG}$. Therefore, many scattering events are required to equilibrate hot electrons with larger energy to the (cold) lattice. In this case, if we take disorder into account, phonons with momenta larger than $2k_F$ can scatter in a three-body collision termed as super-collision[38]. This mechanism is shown to increase the phonon-cooling rate at $T_{ph} > T_{BG}$ and dominates over conventional acoustic phonon cooling for temperatures $T_{ph} > T^*$ where $T_{ph}$ is the phonon temperature and [38] $T^* = \left(\dfrac{\pi}{6\zeta(3)} k_F l\right)^{1/2} T_{BG}$. It was shown by Song et al. that super-collision cooling power per area is given by[38],

$$P = \Sigma(T_{el}^3 - T_{ph}^3); \Sigma = 9.62 \dfrac{g^2 N^2(E_F) k_B^3}{\hbar k_F l} \tag{16}$$

Here, $N(E_F)$ is the density of states per spin and per valley degeneracy and $g = D/\sqrt{2\rho s^2}$ is the deformation potential coupling. The calculation is carried out using Fermi's golden rule listed above by considering impurity scattering before and/or after phonon scattering. This mechanism was experimentally verified using Johnson noise technique to measure the electron temperature[50]. $T_{BG}$ can be tuned above and below $T_{ph}$ by tuning the Fermi energy. Clean limit, low-temperature behavior of the power emitted into phonons, $P \propto T^4$ and super-collision $P \propto T^3$ can be observed in the same device (see Figure 7). Supercollisions are most clearly observed near the charge neutrality point which tunes $T_{BG}$ to a small value of few Kelvin. Since, in the experiment high DC power is applied, $T_{ph}$ can be tuned to be higher than $T_{BG}$. Due to a large potential fluctuation of ~ 65meV at CNP, the highly doped regime of $\mu >> k_B T_e$ is retained, required for super-collision realization. Experiments investigating cooling rates with photocurrent generation have also confirmed the super-collision regime[51].

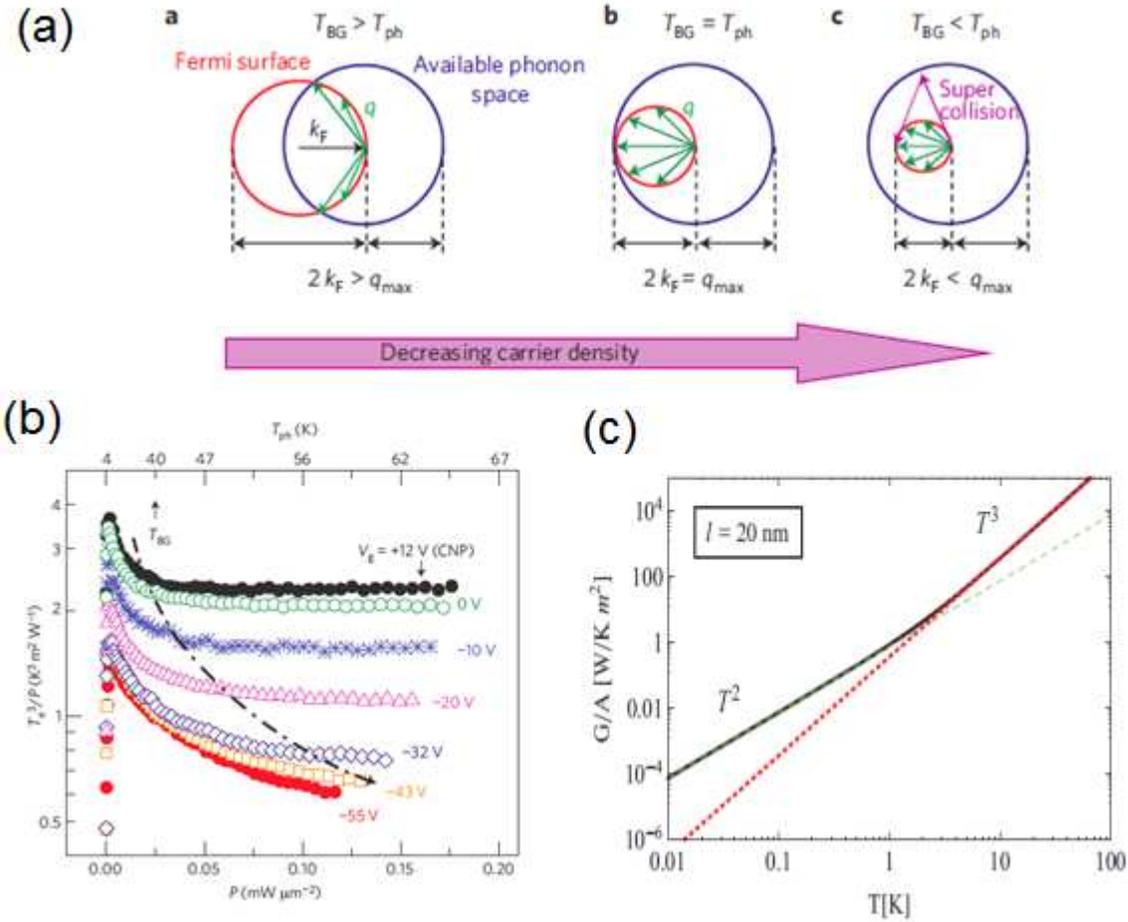

Figure 7. Disorder assisted electron-phonon scattering. (a) Phonon space available for scattering relative to Fermi surface in different temperature regimes as well as in the case of supercollision. From Betz el al.[6]. (b) Experimental demonstration of low temperature $T^4$ cooling power at $T_{ph}<T_{BG}$ and supercollision at $T_{ph}>T_{BG}$, from Betz el al.[6]. At $T_{ph} < T_{BG}$, the cooling power follows $P \sim T_e^4$, and the plotted data on $T_e^3/P$ vs. $P$ follows a $1/T_e$ vs. $T_e^4$ relation. At $T_{ph}>T_{BG}$, $P \sim T_e^3$ and hence $T_e^3/P$ becomes constant. (c) Effect of strong disorder on phonon thermal conductance in the low temperature regime (T $\ll$ $T_{BG}$). The green line shows thermal conductance per area for $T \ll T_{dis}$ and the red line shows the clean-limit, high temperature behavior.

In the case of strong disorder (short mean free path), it is possible that the phonon wavelength becomes longer than electronic mean free path. A new temperature scale then comes into picture below which disorder effects become important[42]

$$k_B T_{dis} = hs/l \qquad (17)$$

where $l$ is the mean free path. If we assume $k_F l \gg 1$, $T_{dis}$ is necessarily well below $T_{BG}$. In this case of high impurity level, scattering calculations based on the golden rule are not useful and the Keldysh formalism was employed to obtain cooling power per area for deformation coupling given as,

$$P = \frac{2\varsigma(3)}{\pi^2} D^2 \frac{E_F}{\hbar^4 \rho_m s^2 v_F^3 l} (k_B T)^3 \qquad (18)$$

In this study, effects of disorder were included to show that when screening is considered, the deformation coupling induced scattering rate is reduced and vector potential coupling related scattering, which arises from the Dirac Hamiltonian as a gauge field, is enhanced. Even though, this is the case for the screening effect, couplings for the unscreened deformation potential are largest and other effects remain less important. Below we summarize some of the main results for the temperature dependence discussed so far in different regimes.

| | Clean limit $T<T_{BG}$, $T_e<\mu$ | Disordered limit $T>T_{BG}$, $T_e<\mu$ | Disordered limit $T<T_{BG}$, $T_e<\mu$ |
|---|---|---|---|
| Cooling Power $P$ (per Area) | $\Sigma(T_e^4 - T_{ph}^4)$ $\Sigma = \frac{\pi^2 D^2 \|\mu\| k_B^4}{15 \rho_m \hbar^5 v_F^3 s^3}$ | $\alpha(T_{el}^3 - T_{ph}^3)$ $\alpha = 9.62 \frac{g^2 N^2(\mu) k_B^3}{\hbar k_F l}$ | $\frac{2\varsigma(3)}{\pi^2} D^2 \frac{\mu}{\hbar^4 \rho_m s^2 v_F^3 l} (k_B T)^3$ |

Table 3. phonon cooling power in the clean and disordered limits in graphene.

## II. Graphene-based bolometers: Experimental Approaches

The general approach of graphene based bolometers is to use graphene as the radiation absorber and detect either the photon energy (phonon counting mode) or the radiation power (power detection mode) by measuring the electron temperature rise due to the incoming radiation. In this section we will review the experimental approaches for realizing graphene based bolometers.

Much of the current work focuses on techniques for measuring the electron temperature in the graphene absorber through transport techniques. A major difficulty in the most straightforward resistive readout scheme for detecting electron temperature in single layer graphene is that the resistivity depends only very weakly on temperature in the low temperature regime. This is due to the weak electron-phonon coupling. This also imposes a challenge for reading the electron temperature through the graphene resistance. For example, at 4.2K, the electron-phonon scattering contributes $\ll 1\Omega$ out of $\sim K\Omega$ for a technically relevant carrier density of n$\sim 10^{12}$ cm$^{-2}$. The corresponding responsivity from measuring $R(T)$ is too small for a practical application.

To address this challenge, various measurements of the electron temperature in the graphene device have been developed, as summarized in Table 3 (in all cases here, the heating was due to a DC or low frequency current, not THz photons). In this section we will discuss these experimental approaches categorized by the readout techniques.

| Reference | # Layers, Contact metal (N = non SC) | Method of temp. readout | $R_{square}(\Omega)$ Temp. range | Dominant Cooling | Phonon mechanism $P = A\alpha\left(T_e^{\delta} - T_{ph}^{\delta}\right)$ ($\alpha=\Sigma$ if p=4) |
|---|---|---|---|---|---|
| Fong et. al[10] | 1, N | Johnson Noise | ≈10 kΩ 10 K | El-phonon | $\delta \approx 4$ $\Sigma$ = 70 mW/m$^2$K$^4$ |
|  | 1, N |  | < 1 K | Diffusion | Matches WF |
| Betz et. al[6] | 1, N | Johnson noise | 1.6-10kΩ ≤ 100K | El-phonon | $\delta \approx 4$ $\Sigma$ = 0.4-2 mW/m$^2$K$^4$ |
|  |  |  | ≥ 4K | Diffusion |  |
| Yan et. al[5] | 2, N | R(T) | 20 -40 kΩ 5-10 K | El-phonon |  |
| Borzenets et.al[12] | 1, SC (Pb) | Hysteresis of I$_c$ | 0.04-1 K | El-phonon | $\delta \approx 3$ At 1 K, $\alpha$ ~60 mW/m$^2$ |
| Vora et.al[9] | 1, SC tunnel (Al/TiO$_x$) | R(T) below T$_c$ | ≈ 2 kΩ 0.15 to 10 K | El-phonon | $\delta \approx 3$ At 4 K, $\alpha$ ~100 mW/m$^2$ |
| McKitterick et.al[7,52] | 1, SC tunnel (NbN/TiO$_x$) | Johnson noise | ≈ 1 kΩ T < 10K | Reduced G, T<T$_c$ |  |
| Voutilainene et.al[11] | 1, SC (thin Ti/Al) | I$_c$(T) | ≈ 1 kΩ 80mK -1K | Tested energy diffusion |  |

Table 4. Comparison of Experimental Reports for Graphene Thermal Properties

## 3.1 Semiconducting bilayer graphene

Bilayer graphene provides a partial solution to the temperature independence of the resistance of monolayer graphene. When the top and bottom layers of a bilayer graphene are doped (electrically or chemically) to opposite carriers bands, lattice inversion symmetry is broken and a semiconducting gap opens at the Dirac point[53]. When tuning the Fermi energy into the semiconducting gap using a dual gate, the bilayer graphene can show temperature dependent resistivity which can be used for measuring the electron temperature.

Yan et.al demonstrated a dual gated bilayer graphene hot electron bolometer (DGBLG HEB)[5]. The device consists of bilayer graphene tuned simultaneously by bottom and top gates to allow generation of an energy gap and tuning of the Fermi level into the energy gap. The device was measured using a 4 probe setup for resistance and placed under illumination from laser light of 10.6 µm wavelength. By comparing the optical and transport measurements, Yan et.al identified the photoresponse to be predominantly bolometric, with the voltage readout signal $\Delta V \sim \frac{dR}{dT} \Delta T$ : at low temperatures, the optical absorption response and the response from electrical Joule heating showed comparable magnitude, indicating the energy from the incoming photons heats up the electrons in graphene to an effective temperature. An interesting observation was the absence of the optical phonon contribution even at large photon energy. This is because the electron-electron scattering is much faster than the electron-optical phonon scattering, allowing the electrons to quickly thermalize among themselves to a temperature where optical phonon emission is weak.

Based on the measured dependence of the electron temperature on heating power, a DGBLG HEB demonstrates an electron-phonon thermal conductance of $G \propto T^3$, which qualitatively agrees with the theoretical expectation for phonon cooling in the clean limit.

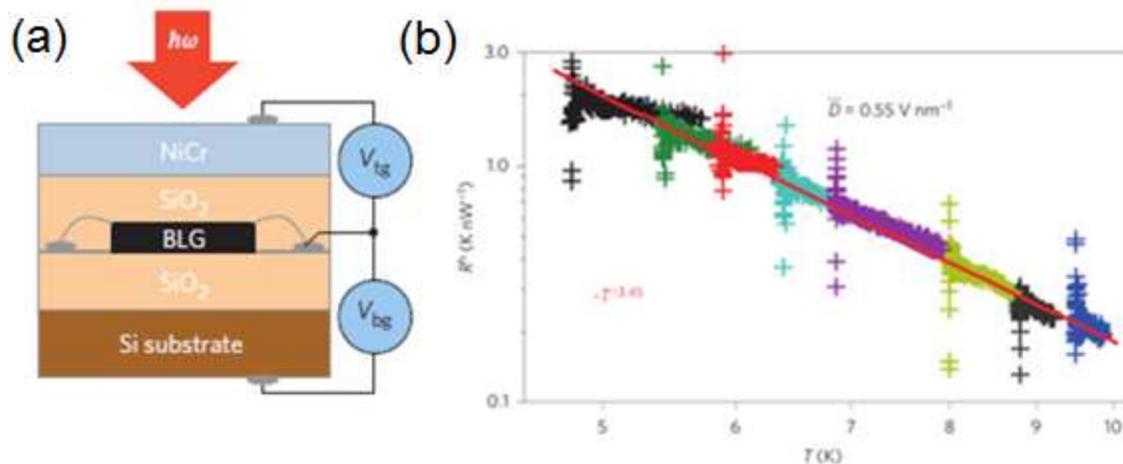

Figure 8. Bilayer graphene bolometer, from ref[5]. (a) Schematics of the device structure. (b) The log-log plot of the measured thermal resistance, $G^{-1}$, which is fitted to a power-law temperature dependence (red straight line) an exponent of ~3.45.

The response time of the DGBLG HEB was measured through a pump-probe technique, utilizing a nonlinear photoresponse. A time constant of 0.25ns at 4.55K and 0.1ns at 10K was observed. Using the measured value of *G* at 5K, the thermal-fluctuation-limited noise equivalent power $NEP = \sqrt{4k_B T^2 G} \sim 2.6 \times 10^{-16} W/\sqrt{Hz}$ was calculated. Unfortunately the Johnson–Nyquist noise of the graphene would be much larger, and limits the noise equivalent power to be $3.3 \times 10^{-14} W/\sqrt{Hz}$. In general one needs $\frac{dR}{dT} \gg \frac{R}{T}$ to be able to achieve the thermal fluctuation limited NEP.

A major challenge in making practical DGBLG HEB is the impedance mismatch between the highly resistive semiconducting absorber (~10-100 kΩ) and free space $\frac{2h}{\alpha e^2} \sim 377\Omega$. A typical planar antenna would prefer a detector impedance ~100Ω for good photon coupling efficiency. The large resistance of bilayer graphene results in an extremely low coupling efficiency. Solving this problem would require the absorber to have much lower resistance compared to the semiconducting bilayer graphene. This appears to be a fundamental challenge.

**3.2 Johnson Noise Thermometry**

The temperature of the graphene detector can be 'read out' by measuring the emission of noise at RF or microwave frequencies. This kind of thermometry is used in low temperature physics labs, and its origins in the Dicke Radiometer[54] are well known. For a source resistor R at temperature T, the average power emitted by the source and absorbed by a matched resistive load is

$$P_J = k_B T B \tag{19}$$

with B being the coupled RF bandwidth. We will treat the case of a matched load. This thermometry approach is desired because the resistance of the graphene with metallic (non-superconducting) contacts is essentially temperature independent. This is true if we choose graphene with an electron density n = $10^{12}$ /cm$^2$ that provides low resistance. A reasonable impedance match to the planar antenna and to the RF amplifier can be achieved with a device resistances of 50 to ≈100 ohms. It is possible to test aspects of detector physics with graphene with much higher resistance, and use resonant coupling to transform that impedance down to match the 50 ohm impedance of the readout amplifier[10]. However, resonant impedance transformation to match the antenna impedance at the THz photon frequency is almost always impractical for a practical, broadband detector. Thus, a low detector resistance at the THz frequency is needed for efficient THz coupling. Up to frequencies of at least a few THz, graphene is resistive and follows the Drude model[55]. Thus, the dc, rf/microwave and THz impedances are approximately the same value and are resistive.

The Johnson noise readout presumes that the electron temperature is in the 'high temperature' limit: for a measurement using an RF/microwave amplifier with center frequency of 1 GHz, the electron temperature should be T > hf/$k_B$ = 50 mK. $P_J$ is linear in temperature, so a measurement of power when the temperature is changing will reflect an average of T during that interval. An important limitation of this readout method is that Eq. 19 only predicts the <u>average</u> power. In a finite time interval τ, the apparent temperature read out will have a FWHM variation for repeated measurements given by the width (FWHM)

$$\delta T_{readout} = 2.3(T + T_A)/\sqrt{B\tau} \tag{20}$$

$T_A$ is the noise temperature of the amplifier[56]. We assume the amplifier does not interact with the detector other than to provide an impedance matched load which fully absorbs all the emitted power. For the single-photon detector we choose the time interval to be equal to the thermal time constant $\tau$ of the photon detector (Table 1). To compute the NEP$_{readout}$ for a power detector, we choose $\tau = 0.5$ sec, corresponding to a 1 Hz noise readout bandwidth. For single-photon detectors that operate in the linear range,

$$\delta E_{readout} = C\delta T_{readout} \tag{21}$$

The total energy width is $\delta E_{tot} = \left(\delta E_{int}^2 + \delta E_{readout}^2\right)^{1/2}$. The Resolving Power is given as $R = E/\delta E_{tot}$. Here and in the following we use the energy width that is the FWHM.

For detectors like Design A with non-linear energy response, one cannot use Eq. 21 to compute the energy Resolving Power. Instead, one needs to calibrate $\delta T_{readout}$ and $\delta T_{int} = \delta E_{int}/C$ against $\Delta T$. These widths must be evaluated at the elevated temperature, ($T_o + \Delta T$), with $\Delta T$ being the average temperature increase during the pulse.[7] The total temperature width is

$$\delta T_{tot} = \left(\delta T_{int}^2 + \delta T_{readout}^2\right)^{1/2} \tag{22}$$

The total resolving power in Table 1 for the non-linear case is

$$R = \Delta T/\delta T_{tot} \tag{23}$$

To maximize the Resolving Power, one would desire a large readout bandwidth and large $\tau$ in Eq. 20. Unfortunately, these are in conflict. The cooling by emission of Johnson noise is characterized by a thermal conductance $G_{photon} = dP_J/dT = k_B B$. Thus, in Eq. 20, $\tau$ decreases when B is increased. Optimizing detector sensitivity requires optimization within these conflicting demands. Moreover, for the most sensitive amplifiers, the amplifier noise temperature approaches the quantum limit, $T_Q \approx hf/k_B$, with f the center frequency of the amplified frequency band. This imposes another constraint for optimizing Eq. 20.

We presented in Table 1 specific predictions for the performance of a graphene bolometer as a single photon detector. The performance listed for a graphene photon detector of Design A in Table 1 is promising. We plot in Figure 9a the schematic response ΔT of Design A as a function of photon energy. For the case of large heating with Design A, ΔT is the average temperature increase during the pulse. In Figure 9b we show the response of hypothetical device that has linear response with a fixed, small energy width $\delta E_{fwhm} = 0.3 E_o$ independent of energy. Each boundary of the hatched areas in Figs. 9a and 9b is defined by the rms energy width. (The FWHM energy width = 2.35 $\delta E_{rms}$ ) We consider illuminating each device with photons of energy $E_o$ and $2E_o$. We plot in a histogram the device response of each device, proportional to ΔT, for an ensemble of absorbed photons of energy $E_o$ and $2E_o$, and also the result of sampling the baseline when there are no photons, the 'zero-photon' peak; see Figure 9c and 9d. Design B, the large area graphene power detector of Table 1, also has a linear response but with a much larger fractional energy width; at $E_o = 1$ THz; $\delta E_{FWHM} = 2E_o$. In the lower panel we plot with dashed lines the outline of the histogram of counts one would measure with design B, with $E_o$ for a 1 THz photon, and also for sampling the baseline. It is evident that Design B would not be useful for counting single photons. Indeed, it was not designed for that application.

In Figure 9a the total FWHM energy width δE increases with increasing photon energy, because the temperature during the pulse is significantly higher for larger photon energy. For example, with design A and a 1-THz photon, the average temperature is 0.6 K during the pulse.[7] This increases both $\delta E_{int}$ and $\delta E_{readout}$ to be larger than with no photons. We plot in Fig. 9c only the zero-photon and one-photon response for design A. For the one photon response, $R = 2.2$, while the zero photon histogram has much narrower width; these results allow good resolution of the 'single-photon' peak, and allow photon counting at rates up to ≈ $10^5$/s with this design.

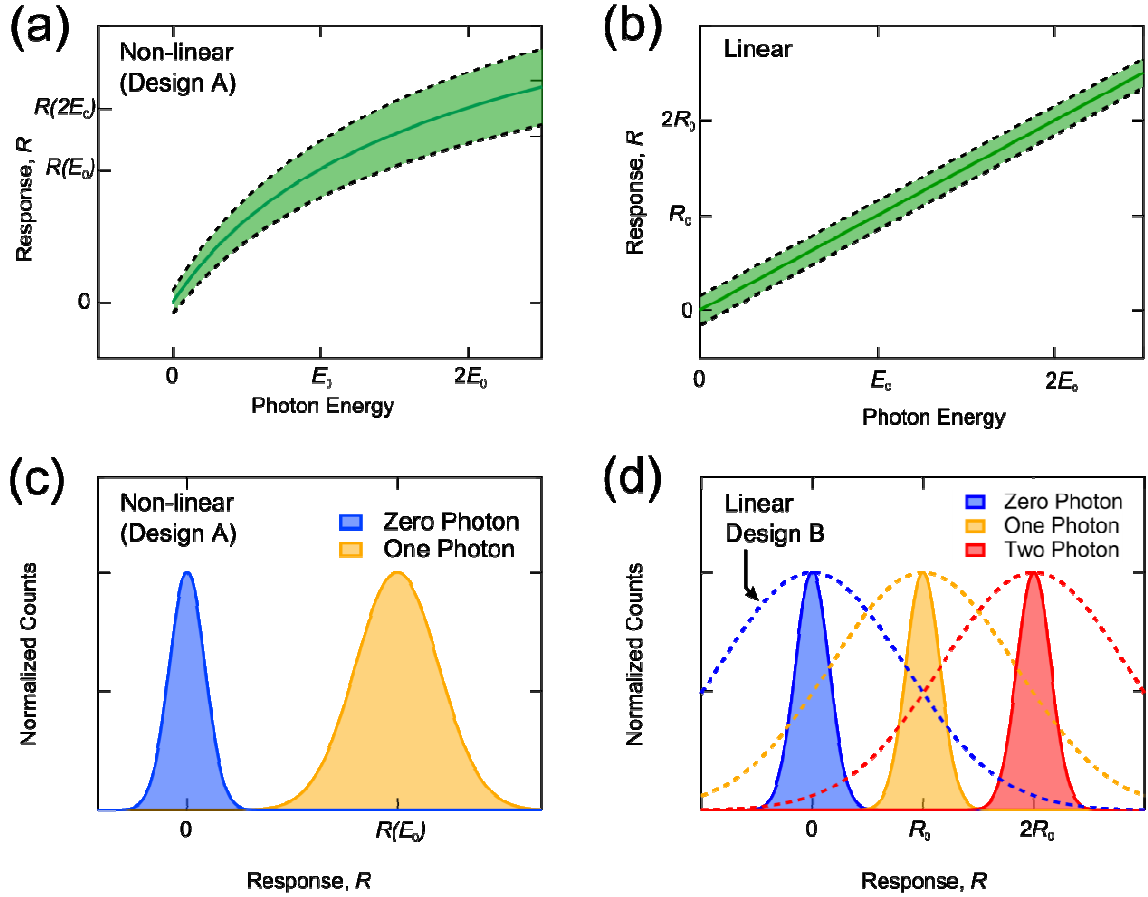

Figure 9. Response of graphene single-photon detector as a function of incident photon energy for a detector that behaves non-linearly (a) and a detector with a linear response (b). The non-linear detector uses the parameters for design A as described in Table I and the (hypothetical) linear detector of panel (b) assumes a $\delta E_{rms} = 0.15 E_o$. Panels (c) and (d) show shaded histograms which correspond to the non-linear and linear curve of panels (a) and (b), respectively. The dotted lines in (d) show the histogram outline for a detector with much larger $\delta E$, corresponding to design B of Table I, for which $\delta E_{fwhm} = 2\, E_o$.

The photon counting mode of Design A is much like that of a photomultiplier or avalanche photodiode. It is useful for weak signals, with average photon number << 1 during the thermal response time, which is 0.5 μsec.[7] For the particular response of Design A, one cannot cleanly distinguish the two-phonon response from that of one photon. With the relatively large energy width, Design A does not, by itself, accomplish spectroscopy. However, with a cold, tunable narrow-band filter ahead of the bolometer, single-photon spectroscopy can be accomplished by tuning the filter.

There have been no tests of a graphene THz single-photon detector. However, studies of the electron-phonon cooling and diffusion cooling of graphene devices have been carried out above 2K using Johnson noise measurements of the electron temperature to allow prediction of future detector performance We describe these next.

**Experiments employing Johnson noise readout**

Fong et.al experimentally studied graphene with Ti/Au contacts at low temperatures down to 2K using Johnson noise emission at microwave frequencies[10]. To avoid strong microwave losses, the devices were fabricated on highly resistive substrates. The device had a high resistance of ~10kΩ. For tuning the Fermi energy of graphene, a top gate was employed. The device showed a mobility of approximately 3500 cm$^2$/Vs at low temperatures, corresponding to a mean-free path of about 20 nm. The charge-carrier density at the CNP is approximately 2 x10$^{11}$cm$^{-2}$, which is estimated from the width of the resistance maximum around the charge neutrality point. In this region, electron-hole puddles are likely formed[57]. To match the high impedance of the device with that of the RF components, a LC network was used which resonates at 1.161 GHz with a bandwidth of 80 MHz.

The device was measured while applying a DC current for Joule heating. From the applied heating power and the corresponding electron temperature was measured from the Johnson noise power, Fong et.al calculated the thermal conductance, $G_{th} = dP/dT_e$, which was observed to follow $G_{th} = (\delta+1)\Sigma A T^\delta$ with $\Sigma \sim 0.07 W/m^2 K^3$ and $\delta = 2.7 \pm 0.3 \sim 3$, respectively. To probe the response time of the devices, Fong et.al applied a high frequency heating current:

$P = I_{heat}^2 R(1 + \cos(2\omega_{heat} t))/2$. And then applied a modulation tone of $\omega_{mod} = \omega_{heat} - 1kHz$. When $2\omega_{heat}\tau < 1$, $T_e$ is oscillatory and $2\omega_{heat}\tau > 1$ $T_e$ becomes constant, hence 1kHz beat decreases. At T~5K, the measurement yielded a time constant of $\sim 1/4\pi\omega_{heat} \sim 68ps$.

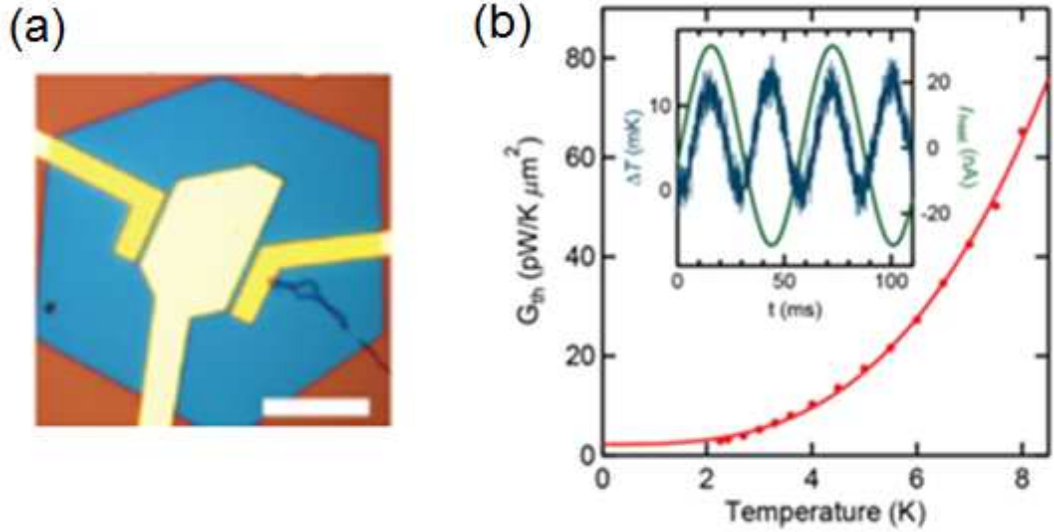

Figure 10. Graphene Johnson noise bolometer, taken from ref[10]. (a) Optical image of the device (top-gated FET with blue, hexagon-shaped gate dielectric). (b) the measured temperature dependence of the thermal conductance, which is fitted to a power-law dependence with power factor of ~2.7. The inset shows the Johnson noise response to a sinusoidal heating current.

Work by Betz et.al studied the dependence of phonon cooling on the quality of graphene, using Johnson noise thermometry[6]. Here the devices with different graphene mobility were fabricated with graphene on BN substrate and CVD graphene on SiO$_2$. The electron temperature was measured from the current noise spectral density $S_I = 4k_B T_e / R$. This was varied through DC Joule heating. Betz et al. found the combined cooling from diffusion and electron-phonon scattering. At sufficiently high bias, it was found that $T_e \propto \sqrt{V}$ (here V is the applied voltage), indicating a cooling power, which follows $P \sim T_e^4 - T_{ph}^4$, a signature of 2D acoustic phonons. At low bias on the other hand, $T_e \propto V$ behavior was observed which corresponds to a cooling power $P \sim T_e^2 - T_{ph}^2$, expected for heat conduction to the contacts following the Wiedemann-Franz law[20]. Studies by Fong et al., above 2K, used a much higher sample resistance, so the diffusion cooling to the contacts was not significant there.

In the phonon cooling regime, Betz et.al observed that the coupling constant increases with increasing carrier concentration. Compared to the theoretical expectation (taking the deformation potential $D \sim 10 eV$) of $\Sigma \sim 10\sqrt{n_s[cm^{-2}]/10^{12}} mW/m^2 K^4$, the observed coupling constant was significantly lower: at $n_s \sim 10^{12} cm^{-2}$, $\Sigma \leq 2 mW/m^2 K^4$ for graphene on BN with a mobility of ~3000 cm$^2$/Vs, and $\Sigma \approx 0.42 mW/m^2 K^4$ for CVD graphene with a mobility of ~350cm$^2$/Vs. Betz et.al attributed the discrepancy to be due to the strong disorder present in the devices, in particular because of the smaller value of coupling constant observed in the low-quality CVD graphene device.

### 3.3 Graphene-superconducting junctions

In graphene-superconductor junctions, the electron temperature in graphene can be obtained from the resistance of the devices. Two types of device configurations have been investigated.

**Graphene-superconductor Josephson weak links.**

We first consider graphene with superconductor contacts that have no tunnel barrier at the interface - transparent contacts[22]. Such graphene devices have several obvious advantages for bolometer applications. First, when biased into a resistive state the devices have low impedance which can be relatively easily matched with external RF/microwave and THz circuits. Second, the superconducting leads prevent hot electrons from diffusing out of the graphene absorber through the Andreev reflection process[58]. On the other hand, there also exist some challenges for these devices to be applicable for practical detectors. To have a voltage to read out, graphene with such Josephson contacts needed to be biased into the finite voltage state, which in turn creates significant self-heating. The IV curve of a device is usually hysteric, due to heating in the finite voltage state[22]. Also the Josephson dynamics can cause significant extra noise and complexity in analyzing the thermal response of these devices[22].

Using the expected hysteretic thermal response, graphene Josephson weak links have been used to study the cooling mechanisms of the hot electrons in graphene. In these studies, the electron temperatures of the devices were deduced through the magnitude of the supercurrent. Such supercurrent measurements do yield reliable measurement of the electron temperature in the strongly self-heated regime, but the large heating would likely preclude their use in a detector operating at low temperatures, e.g., 0.1K.

Borzenets et.al fabricated multiple Josephson weak links on a single graphene sample[12] (see Figure 11). The leads of the device were designed so that the electrical capacitance between the pads coupled through the conducting substrate back gate is small. Thus, the junctions are overdamped[22]. This avoids the hysteresis due to the electrical capacitance that would occur in the underdamped Josephson junction case. The measured difference between switching and retrapping currents was identified to be due to self-heating. Therefore, the switching current can be used as an electron temperature thermometer with Joule heating. Another pairs of leads were chosen to be biased at finite voltage for Joule heating. These leads were spaced far apart so they exhibited a finite resistance at low current. Since electron-electron scattering has a much shorter time scale compare to electron-phonon scattering, with Joule heated electrons are considered to be well thermalized at an elevated temperature throughout the whole graphene sample.

The superconducting contacts thermally isolate the graphene crystal from the leads[22]. In addition, since the work focused on the electron-phonon thermal conductance, a large area graphene sample was used and the measured thermal conductance (which is much larger than what is estimated from the Wiedemann-Franz law by diffusion into the contacts) is dominated by $G_{e\text{-}ph}$. In measuring the cooling power, the graphene was Joule heated. The electron temperature was measured using the Josephson current. The measurements were taken within a base temperature range of ~50 – 700 mK. In this temperature range, different from previous results, this work measured a $T^3$ cooling power of $P = 6 \times 10^{-12} \frac{W}{K^3} T^3$, at a graphene area of $A \sim 100 \mu m^2$. Borzenets et al. compared to the theoretical prediction for acoustic phonon scattering in graphene in the clean-limit[45], and suggested that the observed $T^3$ temperature dependence is because at low temperatures the wavelength of the emitted phonon $hs/k_B T$ becomes longer than another length scale (such as the electron mean free path or spacing between the electrodes[12]). They argued that this imposes a cutoff on the wavelength of the emitted phonons.

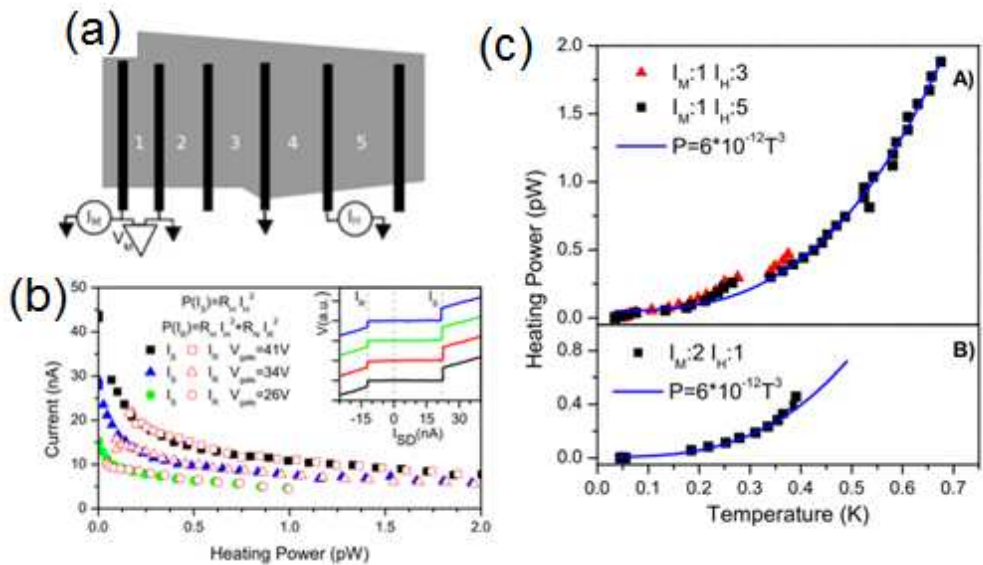

Figure 11. Study of electron-phonon cooling with graphene Josepshon weak links, from ref[12]. (a) Graphene Josephson device for studying phonon cooling. The device has 5 Pb-graphene-Pb junctions on the same piece of graphene. One of the SGS junctions (junction 1 here) can be used as a thermometer and another junction (section 4 here) as a heater. (b) Switching and retrapping currents versus the total heating power at different gate voltages. With sufficient heating the switching and the retrapping currents fall on top of each other. Inset: IV curves under successive current sweeps. The switching and retrapping currents showed negligible fluctuation, indicating that the hysteresis is of a different origin compared to underdamped junctions. (c) heating power vs. electron temperature. The data for *P* can be fit to a temperature dependent power law with exponent ~3. Different sets of data points correspond to different thermometer-heater combinations.

Superconducting contacts with a transparent interface had been previously studied for ultrasensitive bolometers, using a superconducting detector element S' with a lower $T_c$. A good example of such bolometers is the S-S'-S junction[17,18,23,59]. The outer superconducting electrodes, Nb, have larger $T_c$ to confine the hot electrons in the inner Ti S' channel through Andreev reflection. The temperature of the S' channel is tuned to its superconducting transition edge so that its resistance can be used as a sensitive thermometer. In the work on graphene with superconducting contacts with highly transparent interfaces, the superconductors do serve the purpose of confining the hot electrons. However, since graphene is not intrinsically superconducting, the electron carriers temperature had to be measured by biasing the junctions above the critical current. This induces very strong heating and therefore is not favorable for bolometer applications.

One possible alternative to avoid large self-heating at finite bias currents is to build a graphene-superconductor junction with graphene long enough so that the supercurrent is suppressed, yet short enough so that there is still a sensitive resistance-temperature dependence. However, this has the potential difficulty of non-linear effects, due to the highly non-ohmic IV characteristic of the device at zero bias[60].

**Graphene-Superconductor tunnel junctions.**

A graphene superconductor tunnel junction bolometer uses superconducting contacts with a tunnel barrier between each contact and graphene, see Figure 12. The electron temperature can be measured using the quasiparticle tunneling conductance. Below the superconducting transition temperature, the quasiparticle tunneling is suppressed by the presence of the superconducting gap[22]. A strong temperature dependence of the tunneling current and of the electrical tunneling conductance, $G_e = dI/dV$ results from thermal excitation of graphene electrons. The superconducting gap prevents the heated electrons in the absorber from leaking out into the leads if the energy of the warm electrons is lower than the superconducting gap. As a result thermal conductance is reduced, below that of non-superconducting contacts.

The main difference between a tunnel junction bolometer and the other types of graphene bolometers, those with ohmic contacts and those with transparent superconductor contacts, is that the tunnel barrier provides an interface which gives different transparency to DC/low frequency and RF/THz signals (see Figure 12b). For DC signals, the barrier is highly resistive due to the superconducting gap, and therefore hot electron diffusion is strongly suppressed. On the other hand, for RF/THz, the barrier can have very low impedance, shunted by the contact capacitance. For example, for a ~1nm thick $TiO_x$ barrier with a dielectric constant of $\varepsilon \sim 100$, the resulting tunnel junction capacitance is $C_t = \varepsilon \varepsilon_0 \frac{A}{d} \sim 0.9 A[\mu m^2] pF$, and the capacitive impedance is $\frac{1}{2\pi f C_t} \sim \frac{180}{f[GHz] A[\mu m^2]} \Omega$. For a typical contact area of a few $\mu m^2$ and frequency of a few GHz, the capacitive impedance becomes negligibly small compared to the DC resistance of the contacts and the resistance of graphene. This difference in the contact electrical impedance, low vs. high frequency, gives two major advantages for the superconductor tunnel junction bolometer scheme. First, the RF/THz impedance can be made low enough to match the graphene resistance itself with the antenna, allowing high photon coupling efficiency while still keep a low thermal conductance due to suppressed diffusion. Secondly, the RF/microwave readout has a very small voltage drop across the device. As a result, the non-linearity effect in photoresponse can be minimized. Measurement of electron temperature of the graphene superconductor tunnel junction bolometers is also different from that for the superconducting-contact bolometers without the tunnel barrier[12]. Since the current-voltage relation in a superconducting tunnel junction is strongly non-ohmic, the bias voltage $V$ must be small to avoid non-linearity. Hence to achieve certain heating power $P = \frac{V^2}{R}$, the total device resistance $R$ needs to be small. As a result, in characterizing the device sensitivity, Joule heating needs to done by applying a RF/microwave signal which "sees" mainly the small resistance from the graphene absorber. A small DC bias which induces a voltage predominantly across the graphene-superconducting tunneling contacts is used for detecting the electron temperature.

Vora et.al first demonstrated a graphene-superconductor tunnel junction bolometer using graphene-TiO$_x$-Al junctions[9]. Al has a superconducting transition temperature of $T_c = 1.2\ K$. The TiO$_x$ tunnel barrier was obtained by oxidation of thermally evaporated Ti. Ti was chosen for two reasons: the good wetting property of Ti on graphene and the large dielectric constant of TiO$_x$ ($\varepsilon \sim 100$). The devices were designed so that graphene between the tunnel junctions has a large width/length (W/L) ratio and hence low resistance (~100Ω). The DC resistance of the devices is dominated by the tunneling resistance, which was found to be typically between 1-100kΩ.

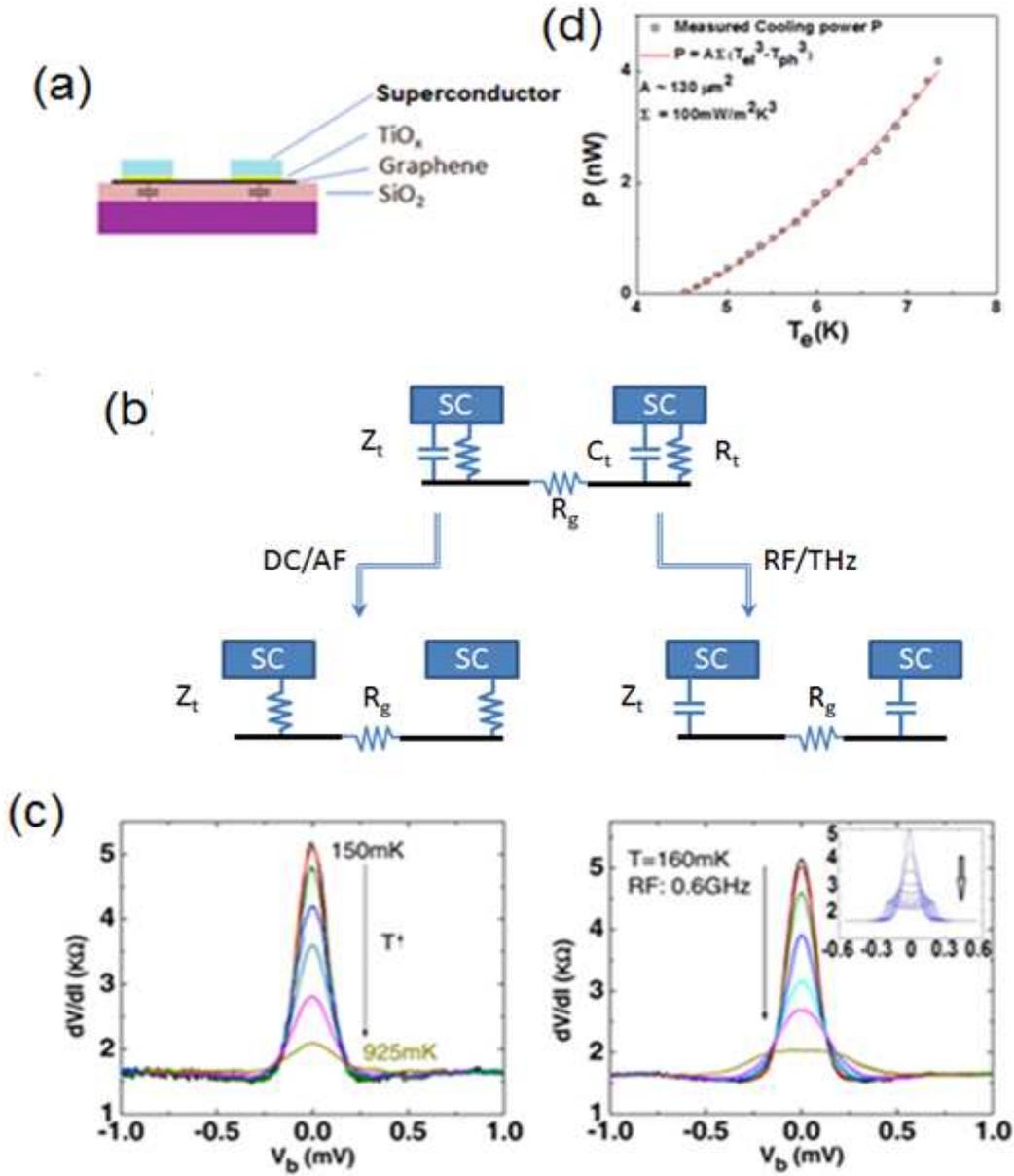

Figure 12. (a) Device structure of a graphene-aluminum tunnel junction bolometer. (b) Equivalent circuit for a graphene-superconductor tunnel junction bolometer. For DC/low frequency signals, the total impedance is dominated by a large tunneling resistance; for RF/THz signals, the junction capacitance $C_t$ shorts out the tunneling resistance $R_t$ and therefore the impedance is predominantly from the graphene absorber: $|Z_t| \approx R_t >> \frac{1}{\omega C_t}$. (c) Comparison between the temperature dependence (left panel) and RF power dependence (right panel) of the differential resistance vs. bias voltage curves. The two conditions yield almost identical results at low temperatures $T<<T_c$. Inset: simulated non-linearity response for the junction. (d) Cooling power vs. temperature measured in a graphene-TiOx-NbN tunnel junction bolometer.

By measuring the temperature dependence of the differential resistance vs. bias voltage in the absence of radiation, and comparing the results with those taken at base temperature but with incoming radiation, Vora et al. found that at low temperatures where the superconducting gap is roughly temperature independent, the RF radiation yields an effect that is identical to the bath temperature increase. This confirmed that the observed RF response in the graphene-TiO$_x$-Al tunnel junctions is indeed bolometric. By correlating the radiation and the bath temperature dependence of the differential resistance in a graphene-superconductor tunnel junction, Vora et al. were able to use the zero bias tunnel resistance as a electron temperature thermometer.

Further development of graphene superconductor tunnel junction bolometers is focused on improving the thermal isolation, by using higher T$_c$ superconducting contacts. With these devices, Vora et al. studied the hot electron cooling mechanism in a temperature range of 4.2-7K. There it was found that[60] the cooling is predominantly from acoustic phonons in the disorder limit, where the cooling power fits[42] $P = A\Sigma(T_e^3 - T_{ph}^3)$. The value of the coupling constant was found to be $\Sigma \sim 100 mW/m^2 K^3$, which is consistent with the theoretical expectation.

The demonstration of bolometric response in graphene-superconductor tunnel junctions [9,60] is a promising approach for building the state-of-the-art power detectors, because it simultaneously allows hot electron confinement and low impedance at high frequencies. A state-of-the-art power detection bolometer should have a low thermal conductance for high sensitivity, operate in the linear regime, and have sufficiently large output response for external amplification. We next discuss the choice of device parameters with these requirements, considering a current-biased graphene superconducting tunnel junction bolometer[9]. First of all, we desire the phonon cooling with $G_{phonon} = \delta A\Sigma T^{\delta-1}$ to be the bottle neck of the thermal conductance, diffusion cooling following Wiedemann-Franz law $G_{diff} = \frac{4LT}{R}$ with should be small: $G_{diff} < G_{phonon}$. Hence we require R, the DC resistance of the device, largely due to the superconducting tunnel junction, to satisfy:

$$R > \frac{4LT}{\delta A\Sigma T^{\delta-1}} \quad (24)$$

Second, for the device to work in the linear regime, we need the temperature increase at absorption of a single phonon to be small:

$$\frac{E}{C} \ll T \quad (25)$$

where $E = hf$ is the photon energy. This requires *large C, hence large graphene area or very large carrier density*.

Finally let us consider the response of the device under radiation with a power $P$, the corresponding voltage change is: $\Delta V = I\frac{dR}{dT}\Delta T = I\frac{dR}{dT}\frac{P}{G}$, $G$ here is the thermal conductance. The current bias is limited by two factors. First of it should not cause significant self-heating. This requires $\frac{I^2 R}{G} \ll T$ hence $I \ll \sqrt{\frac{GT}{R}}$. Second, the induced voltage bias should be much smaller than the superconducting gap for optimized thermal confinement, $I \ll \frac{\Delta}{eR}$. In general, we need $I \ll \min\left(\sqrt{\frac{GT}{R}}, \frac{\Delta}{eR}\right)$.

At the ultimate sensitivity, we need to be able to measure $P = NEP_{int} * \sqrt{B} = \sqrt{4k_B T^2 G B}$. The resulting voltage signal needs to be resolvable above the voltage noise of the amplifier: $\Delta V > S_V \sqrt{B}$ ($S_V$ being the amplifier voltage noise spectral density), hence

$$S_V \leq I \frac{dR}{dT} \frac{P}{G} \frac{1}{\sqrt{B}} << \min\left(\sqrt{\frac{GT}{R}}, \frac{\Delta}{eR}\right) \frac{dR}{dT} T \sqrt{\frac{4k_B}{G}} \qquad (26)$$

Neglecting the diffusion cooling, since eq.24 is satisfied, the requirement on the amplifier noise is:

$$S_V << \min\left(\sqrt{\frac{\delta A \Sigma T^\delta}{R}}, \frac{\Delta}{eR}\right) \frac{dR}{dT} T \sqrt{\frac{4k_B}{\delta A \Sigma T^{\delta-1}}} \qquad (27)$$

For $S_V$ to be practical for a real-life amplifier, we need to make the thermal conductance G sufficiently small, either by using *small area graphene or low carrier density(to reduce Σ)* .

Whether or not a graphene superconducting tunnel junction bolometer is promising depends on the existence of parameters which satisfies requirements Eq.24-26. Consider a graphene THz photon power detector ($E_{ph} \sim 7 \times 10^{-22} J$) working at T = 0.1 K with a carrier density of $10^{12}$ cm$^{-2}$, using $C \sim A[\mu m^2] \times 7 \times 10^{-22} T[K]$ (see Figure 3), we find the area of graphene needs to be $A >> 100 \mu m^2$ for the device to have linear response. Assuming $A = 1000 \mu m^2$ and $G_{phonon} = 4 A \Sigma T^3$ (for clean limit) with[6] $\Sigma \sim 0.5 mW/m^2 K^4$, we find from Eq. 24 the resistance of the device should satisfy $R > 5M\Omega$. We can take the tunneling resistance to be $R(0.1K) = 10 M\Omega$. To avoid self-heating, we choose an excitation current of 1 pA which satisfies $I \ll \min\left(\sqrt{\frac{GT}{R}}, \frac{\Delta}{eR}\right)$. We require an practical amplifier noise of $S_V \sim 4nV/\sqrt{Hz}$, and from Eq. 27 we can obtain $\frac{dR}{dT} > 200 M\Omega/K$. Consider superconducting leads with $T_c \sim 0.6K$, and $R \sim R_0 e^{\frac{\Delta}{k_B T}} = R_0 e^{\frac{1.76 T_c}{T}}$, where the superconducting gap $\Delta \approx 1.76 k_B T_c$ and $R_0 \approx 300\Omega$ is the normal state tunneling resistance. We can estimate $R(0.1K) \approx 10 M\Omega$, and $\frac{dR}{dT}(0.1K) = 1 G\Omega/K$, both satisfy the requirements discussed here. The resulting sensitivity is a noise equivalent power of $NEP = \sqrt{4 k_B T^2 G_{phonon}} = 4\sqrt{k_B T^5 A\Sigma} \sim 5\times 10^{-20} W/\sqrt{Hz}$.

With the above parameters, the device will consume a DC power of $10^{-17}$ W. At a THz photon arrival rate of $10^4$/sec ($7\times 10^{-18} W$), the total heating power is $1.7\times 10^{-17} W$ and the electron temperature will increase by 8.5mK. Correspondingly and resistance drops from 10MΩ down to ~5MΩ. The response is reasonably linear. On the other hand at a THz photon arrival rate of $10^5$/sec ($7\times 10^{-17} W$), the total heating power becomes $8\times 10^{-17} W$ and the electron temperature will increase by 40mK. Correspondingly and resistance drops from 10MΩ down to ~500KΩ. The operation is no longer in the linear response regime.

The results above depend on the details of the parameters. A more quantitative knowledge of the coupling constant $\Sigma$, as well as the temperature dependence of the phonon cooling in the millikelvin temperature regime are crucial for designing the state-of-the-art graphene bolometers, and still require further study.

While promising for power detectors, graphene-superconductor tunnel junctions with resistance readout are not applicable for single-photon detection because of their large resistance that is required to give good thermal isolation. The large resistance significantly narrows the bandwidth. For a junction resistance of $R(0.1K) = 10M\Omega$, with an amplifier capacitance of 10pF or a cable and amplifier capcitance of >100 pF, the *RC* time constant would be 0.1 to 1 msec. This is much too long for reading out single photons. Generally a low device resistance at readout frequencies is needed for single photon detection, and Johnson noise thermometry may be used for readout (see section 3.2).

### III.   Conclusions and future challenges

To achieve the state-of-the-art detectors, extensive research has been carried out on graphene-based bolometers, utilizing graphene's promising properties including small heat capacity, weak electron-phonon coupling, and small resistance. Theoretical efforts focus on understanding the phonon cooling mechanism from acoustic and optical phonon modes, as well as the impact of temperature, doping, and disorder on electron-phonon scattering. Experimental work explored various approaches for measuring the electron temperature and for achieving the phonon-cooling bottleneck.

Based on all these efforts, we can design and estimate the performance of an "ultimate" graphene bolometer. At an operating temperature of 0.1K, with superconducting leads to confine the hot electrons, a graphene bolometer can operate with resistance or Johnson noise readout. With the resistance readout, a graphene-superconductor tunnel junction bolometer may operate as a highly sensitive power detector with the NEP reaching $\sim 5 \times 10^{-20} W/\sqrt{Hz}$. Such a device, with its large low frequency resistance, cannot operate as a single photon detector. With the Johnson noise readout, single photon detector can be achieved when operating in the non-linear regime. In the design for operation in the linear regime, Johnson noise readout allows power detection with a NEP reaching $\sim 1.2 \times 10^{-19} W/\sqrt{Hz}$. The sensitivity is slightly less than that in a graphene-superconductor tunnel junction bolometer, but the response is much more linear.

Testing of a sensitive THz detector is very challenging. One needs to employ a known, very small photon flux of narrow bandwidth with low total count rate. A cold blackbody, $T_{bb} \approx$ 4 K, could be employed with a cold bandpass filter[61]. No leakage at lower frequencies (for which hf ≈ $kT_{bb}$) is allowable. A warm blackbody, with $kT_{bb}/h \geq 1$ THz could be employed with significant cold attenuation and a cold bandpass filter; again no leakage at other frequencies is allowable. A quantum-cascade laser might also be used as a source, again with cold attenuation[23]. No out of band emission is allowable. Last, a THz single photon simulation can be done with using a pulse of rf photons with the same energy. The last two test methods allow triggering of the readout; the last method, a 'fauxton' test system, has been employed with success to test a superconducting detector at equivalent photon energies that are an order of magnitude larger than for a 1 THz photon. In all cases cold frequency selective filters and cold attenuators must be employed.

There are a number of physics/device engineering issues that need to be addressed in future experiments. The absorbed THz photon initially creates a high energy (≈ 4 meV) single electron excitation. A 1-THz photon has energy equivalent to $T_{eff}$ = hf/k = 45 K >> $T_o$. The equilibration to ultimately produce an electron distribution at an elevated temperature can be understood using the electron-phonon and electron-electron interactions. The theoretical predictions below 1 K have not been tested in experiment, and the controlling parameters are not yet well known. The possible loss of energetic electrons by diffusion to the (non-superconducting) contacts, or loss by diffusion over the superconducting energy gap of superconducting contacts, can be modeled for each specific geometry and material choice. The effect of amplifier noise on the detector, in band and out-of-band, also needs to be determined. Cryogenic circulators and filters[23] as used in both astronomy detectors and quantum computing circuits will need to be employed. In short, there is much device engineering that will need to be done.

There are also physics questions that are not yet answered. Chief of these is the strength of the electron-phonon coupling at low energies, below T = 1 K, but also at higher energies, up to the photon energy. This will allow modeling of the electron cooling process.

**Acknowledgement**

Xu Du and Heli Vora acknowledge support from AFOSR-YIP grant FA9550-10-1-0090. Daniel E. Prober and Chris McKitterick acknowledge support from NSF Grant DMR-0907082, an IBM Faculty Grant, and Yale University.